\documentclass[%
 reprint,
 amsmath,amssymb,
 aps,
]{revtex4-1}

\usepackage[usenames,dvipsnames]{color}
\definecolor{purple}{rgb}{0.5,0,0.5}

\usepackage{graphicx}
\usepackage{dcolumn}
\usepackage{bm}
\usepackage{tikz} 
\usetikzlibrary{external}
\usepackage{ifthen}
\usetikzlibrary{calc,positioning,decorations.pathmorphing}
\usepackage[T1]{fontenc}

\usepackage{caption}
\usepackage{dcolumn}
\usepackage{bm}

\begin{document}


\title{Gapless Topological Order, Gravity, and Black Holes}
\author{Alex Rasmussen}
\thanks{These two authors contributed equally.}
\affiliation{Department of Physics, University of California at Santa Barbara, Santa Barbara, CA 92092, USA.}
\author{Adam S. Jermyn}
\thanks{These two authors contributed equally.}
\affiliation{Institute of Astronomy, University of Cambridge, Cambridge CB3 0HA, UK.}
\date{\today}

\begin{abstract}

In this work we demonstrate that linearized gravity exhibits gapless topological order with an extensive ground state degeneracy.
This phenomenon is closely related both to the topological order of the pyrochlore $U(1)$ spin liquid and to recent work by Hawking, et al, who used the soft photon and graviton theorems to demonstrate that the vacuum in linearized gravity is not unique.
We first consider lattice models whose low-energy behavior are described by electromagnetism and linearized gravity, and then argue that the topological nature of these models carries over into the continuum.
We demonstrate that these models can have many ground states without making assumptions about the topology of spacetime or about the high-energy nature of the theory, and show that the infinite family of symmetries described by Hawking et al are simply the different topological sectors.
We argue that in this context black holes appear as topological defects in the infrared (IR) theory, and that this suggests a potential approach to understanding both the firewall paradox and information encoding in gravitational theories.  
Finally, we use insights from the soft boson theorems to make connections between deconfined gauge theories with continuous gauge groups and gapless topological order.

\end{abstract}

\maketitle


\section{Introduction}

The black hole information paradox, which calls into question the fate of information falling into a black hole, has led to considerable work on the entanglement structure of quantum gravity theories.
Central to this paradox is the statement that black holes have no hair, which is known to hold classically and was initially thought to hold quantum mechanically as well \cite{PhysRevD.14.2460}.
However, recent work has shown that both the electromagnetic field and the gravitational field contain ``soft'' hair \cite{1601.00921}.
This hair comes in the form of soft (zero-energy) bosons which have long been known to exist in the zero-$k$ limit of these theories \cite{PhysRev.140.B516}.

A key step in the identification of soft bosons with information-carrying soft hair is finding the corresponding large-scale time-dependent symmetry, as classical electromagnetic and gravitational theories obey the no-hair theorem in the steady state \cite{PhysRev.164.1776}.
In the case of gravitation the classical symmetry group is known to be that of Bondi, Metzner, and Sachs, and the corresponding electromagnetic symmetry is similar in structure \cite{PhysRev.128.2851,1601.00921}.

In addition, the soft photon and graviton theorems have played an important role in the relation between symmetries and quantum memories.
This results in the so-called ``triangle'' that relates the soft boson theorems to large gauge symmetries and memories, with deep connections to the Ward identities for those gauge theories\cite{1502.07644, 1506.02906, He2015, PhysRevLett.116.031602, 1505.00716}

Of particular interest is the connection between these ideas and the notion topoligcal order.
Ordinarily, topological order manifests by the existence of global modes which `wrap' around the system and which are only accessible by means of gapped excitations.
In this way such topological modes encode protected quantum information.
Recently it has come to light that gauge theories with similar structure to electromagnetism, gravitation, and higher-order equivalents generically exhibit a peculiar variant of this phenomenon\ \cite{Hermele2004, Xu2006, 1601.08235, 0602443, 1604.05329}.
The peculiarity stems from the fact that these gauge theories have a stable, deconfined IR Gaussian fixed point, and thus have exactly gapless gauge bosons in the spectrum.
Nevertheless, their ground states are degenerate on a torus and indistinguishable by local operators.
As such they exhibit protected topological charges which, in contrast to more typical systems, are protected by large-scale gauge symmetries rather than by an energy gap.

In this work we show that these two observations are intimately related: the soft hair which \citet{1601.00921} discovered corresponds to topological zero-modes which live on the boundaries of our low-energy phase of spacetime, be they at infinity or at the horizon of a black hole.
Equivalently, states with different numbers of soft bosons correspond to different topological sectors.
As a result, the degeneracy of the gravitational vacuum is really a reflection of the underlying ``gapless topological order'' of gravitation and electromagnetism.
Notably this result holds even though spacetime at a glance is simply connected, and we show that this is a direct consequence of the metric signature and gapless nature of soft modes.

This work also yields a possible resolution to the firewall paradox of \citet{Almheiri2013}.
Outgoing Hawking radiation can be entangled initially with its infalling counterpart, but upon interaction with the soft sector at the horizon loses this entanglement.
This interaction is required by the correspondence between the soft sector and flux integrals that can be performed around the black hole.
As a result it is equivalence, postulate (4) of \citet{Almheiri2013}, which is violated.
Interestingly this violation is purely quantum mechanical, as it relies on scattering with the soft sector, which cannot be detected except via entanglement measurements.
In this way the classical equivalence principle is preserved.

This paper is organized as follows.  In section II, we review the constructions of lattice QED and lattice linearized gravity.  In section III, we analyze the topological winding procedure and provide a much more concrete description of topological degeneracy in gapless systems.  In section IV, we demonstrate how matter falling into a black hole can be seen as changing topological sectors.  The remainder of the paper analyzes some subtleties of these phenomena and speculates on the implications of the structure of spacetime and the information paradox. 

\section{Gapless Topological Order and Deconfined Gauge Theories}

The first system exhibiting what we call gapless topological order\footnote{Note that this is a distinct phenomenon from ``quasi-topological order'' \cite{bonderson2013}.}  was the $U(1)$ spin liquid on the pyrochlore lattice\cite{Hermele2004}.  The gapless excitation is a ``photon'' for an emergent $U(1)$ gauge symmetry, and the spinons carry electric charges.  The ground state degeneracy was shown to be manifold-dependent, and argued to be stable in the presence of a spinon gap and infinite system size.  However, the ground state degeneracy only closed with the ground state as $1/L$, which have the same energy as the lowest-lying photon states.  Later works\cite{Xu2006, 1601.08235, 0602443, 1604.05329} found similar topological degeneracy in stable gapless phases -- all of which are gauge theories.

As we will see below, these degenerate ground states should instead be identified with the ``soft'' gauge bosons.  The operator that inserts soft bosons will be shown to be the same that moves between degenerate ground states.

Importantly, not all stable gapless systems inherit this structure.  Systems with spontaneously broken 0-form\footnote{I.e. the symmetry acts on point-like objects} symmetries, such as superfluids, lack the gauge structure necessary for constructing the topological sectors.  Systems with gapless matter, such as Weyl semimetals, are also excluded even if there is a gauge structure.  In this second case, the photon may still be stable but the topological sectors will not be protected by the charge gap.

Furthermore systems without the appropriate gauge structure may exhibit power-law splitting and local indistinguishability but have no low-lying modes which are sensitive to the topology of the system.
Thus, for instance, modes which are localized on a scale $L^{1/2}$ may be gapless in this sense and may be locally indistinguishable yet not be global and hence not provide topological charge.

In section III we discuss the idea of spontaneously broken higher-form symmetries, which connect the gauge structure to topological sectors and Wilson lines.  This provides a unified way to understand topological order in gauge theories with both discrete and continuous gauge groups, but may not be sufficiently general to characterize all topologically ordered phases.

To explicitly draw a connection between the deconfined gauge theories and the gravitational ground state, we first review the constructions of two relevant lattice systems - electromagnetism and linearized gravity.  In particular, we stress that these models can be built from local bosonic degrees of freedom on a lattice, and that the corresponding emergent gauge theories exist at exactly stable IR (continuum) fixed points.

By emergent gauge theory, we mean that the theory has a low-energy Hilbert space with local constraints $\hat{Q}(x)$, all of which commute with the Hamiltonian and each other.  For example, this can happen for an easy-axis Heisenberg model on the cubic lattice when typical energies are smaller than the exchange coupling\cite{Hermele2004}.  Physical states \textit{in this reduced Hilbert space}, i.e. that with such low energies, are closed under these operators, which is to say that
\begin{equation}
\hat{Q}(x)\left|\mathrm{Phys}\right\rangle = 0.
\end{equation}
Closure under $\hat{Q}(x)$ generates corresponding local conservation laws.  The model then becomes a gauge theory when we identify the physical low-energy states that differ only by a gauge transformation.  We can then write an effective low-energy field theory in terms of the gauge field for these constraints at the IR fixed point.

\subsection{Electromagnetism}
\label{sec:em}

\begin{figure}
	%
%
			%
%
		%
		%
		%
	\includegraphics{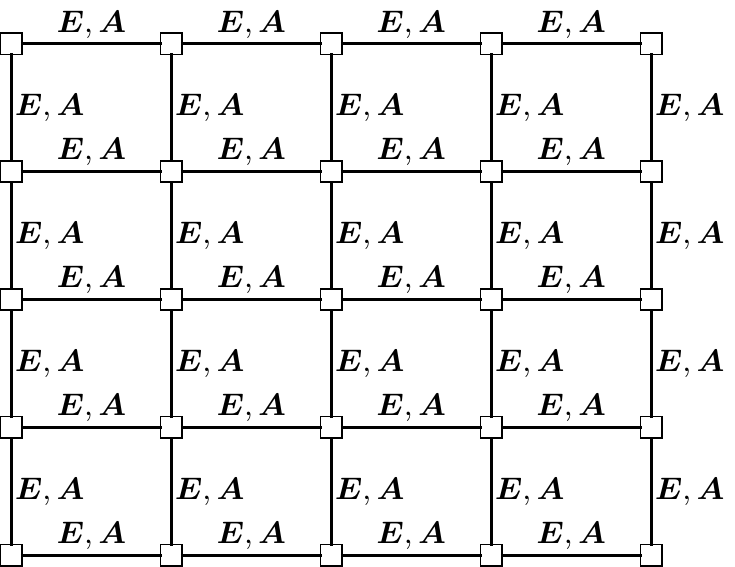}
	\caption{A two-dimensional slice of the lattice version of our system is shown. Only a $5\times 5$ sublattice is shown, but the system may arbitrarily large in each dimension. Likewise the system shown may have open (as shown) or topologically non-trivial boundaries. Note that the conjugate vector fields $\boldsymbol{E}$ and $\boldsymbol{A}$ live on the links in the lattice, and for clarity these are only shown in the open boundary case.}
	\label{fig:lattice}
\end{figure}

First, we review the simplest case of gapless topological order -- ordinary QED in $3+1d$  Following \cite{Hermele2004}, this quantum theory has a compact $U(1)$ gauge group with two canonically conjugate vector fields $\boldsymbol{E}_i \in \mathbb{Z}$ and $\boldsymbol{A}_i \in [0,2\pi)$ (along with the corresponding operators $\hat{E}_i$ and $\hat{A}_i$) that live on the links of a cubic lattice as shown in Figure\ \ref{fig:lattice}.  This model can be derived starting from the Heisenberg model and introducing easy-axis anisotropy.
Because charged excitations are gapped, the low-energy Hilbert space has a local conservation law
\begin{equation}
\label{em_constraint}
\partial_i \hat{E}_i\left|\mathrm{Phys}\right\rangle = 0,
\end{equation}
which just follows from Gauss's law.  By considering the commutator $[\hat{A}_i(x),\hat{E}_j(z)] = i\delta_{ij}\delta(x-z)$ and a local phase rotation $\exp{\int{d^dx \lambda(x)\partial_i\hat{E}_i(x)}}$, we see that $\hat{A}$ is shifted by

\begin{equation}
\label{em_transform}
\hat{A}_i \rightarrow \hat{A}_i + \partial_i \lambda,
\end{equation}
where the derivative operator acts as a finite difference on the lattice.
After taking the spin wave limit\cite{Hermele2004}, we see that the most relevant terms in the low-energy effective Hamiltonian (acting on this reduced Hilbert space) are 
\begin{equation}
\hat{H} = \frac{U}{2}\sum_i \hat{E}_i^2 + K\sum_i \hat{B}_i^2
\end{equation}
where we have defined $\hat{B}_i = \epsilon_{ijk}\partial_j \hat{A}_k$ as the usual curvature, and $U$ and $K$ depend on the microscopic couplings.

In the following sections, it will be useful to think of the local constraint (and its accompanying gauge transformation) as the essential component of the field theory.  It is argued by \citet{Hermele2004} that the IR fixed point defined by this gauge theory is completely stable, and that all other terms are irrelevant in the renormalization group sense.  Thus the actual lattice realization of this theory is not enormously important, provided that this local constraint is enforced in the IR.

Gauge-charged matter in the theory show up as defects of this conservation law.  This follows from the Gauss constraint
\begin{equation}
\left(\partial_i \hat{E}_i - \hat{\rho}\right)\left|\mathrm{Phys}\right\rangle = 0
\end{equation}
which enlarges the original gauge constraint to include charged matter.  We note that the tensor form of $\hat{\rho}$ is determined by the constraint.  Furthermore, the energy gap of the charged matter (i.e., the mass of the spinons) has to be large compared to other scales to enforce the constraint.

A simple, but insufficiently general, argument to demonstrate a topological degeneracy starts by putting the system on the three dimensional torus $T^3$. Then, we create a charge-anticharge pair and propagate them around a non-contractible loop of the torus.  This threads a single electric flux, which can spread out over the whole system uniformly and thus has total energy that goes to zero as $1/L$ (for system size $\sim L$).  Without loss of generality we can assume the winding direction (and thus flux) is perpendicular to a surface $\Sigma$ with normal vector in the $i$ direction.  Then, the topological sector is determined by flux integrals
\begin{equation}
\hat{\Phi}_i = \int_\Sigma {dS_i \hat{E}_i} = \left(\int_\Sigma \star F\right)_i,
\end{equation}
where the index $i$ is not summed over.
These integrals compute the electric flux through a surface $\Sigma$ perpendicular to the components of $E_i$.  Since the charges of $E_i$ are labeled by integers, the fluxes are also integers.  The flux integrals commute with each other and the Hamiltonian 
\begin{equation}
\left[\hat{\Phi}_i,\hat{H}\right] = 0
\end{equation}
Thus, the ground states are labeled by three integers, corresponding to the eigenvalues of these electric flux integrals.  Because $\hat{A}$ is compact, we should also include monopoles in the spectrum on the lattice \cite{Hermele2004}, which have a corresponding interpretation in the continuum \cite{PhysRevLett.116.031602}.  However, this only expands the number of topological sectors by adding three integers corresponding to the magnetic flux winding, and is not essential to our results.

This argument holds when the system lives on $T^3$, but runs into several problems when the system is put on, say, a solid torus (one periodic dimension and two open dimensions).  This difficulty is addressed in Section III.

\subsection{Linearized Gravity}

To discuss gravity as a gauge theory requires leaving Yang-Mills behind.  First, the gravitational gauge group is non-compact due to the four translations, and we must be careful about gauging the local rotations of the frame fields.  Second, the gauge field is no longer an algebra-valued 1-form field, but instead a symmetric 2-tensor.  This means that the gauge-charged matter carries a Lorentz index instead of a color index:

\begin{equation}
\left(\partial_\mu \hat{T}_{\mu\nu} - \hat{\rho}_{\nu}\right)\left|\mathrm{Phys}\right\rangle = 0.
\end{equation}

Since the stress-energy tensor $\hat{T}_{\mu\nu}$ is the generator of translations, we identify the gauge charge as the momentum carried by an excitation.  This identification is valid in the linear regime where gravitons do not couple to one another and hence cannot themselves carry gauge charge.

We want to draw an explicit connection to a lattice model, so first we will do a partial gauge fixing and then a linear approximation.  The first step is to foliate spacetime in a timelike direction using the ADM formalism\cite{rezzolla2008}.  This is a partial gauge fixing of the full gauge group (in particular, we use the synchronous gauge), and the new dynamical variables are the symmetric 2-tensor spatial metric $A_{ij} \in [0,2\pi)$ on each slice and its conjugate $E_{ij} \in \mathbb{Z}$ (the stress tensor), along with the corresponding operators $\hat{A}_{ij}$ and $\hat{E}_{ij}$.  We can then linearize this theory, considering only small fluctuations around the background metric.

The lattice bosonic rotor model we consider \cite{0606100} reproduces these variables with $\hat{E}_{xx}$, $\hat{E}_{yy}$, and $\hat{E}_{zz}$ (along with their conjugate $\hat{A}$) living on each vertex of a cubic lattice, while $\hat{E}_{xy}$ and similar living on the faces.  Then we set up the Hamiltonian to enforce the following constraints in the low-energy:
\begin{subequations}
\label{lin_adm_grav_constraint}
\begin{align}
\partial_i \hat{E}_{ij}\left|\mathrm{Phys}\right\rangle &= 0 \\ 
\left(\delta_{ij}\partial^2 - \partial_i\partial_j\right)\hat{A}_{ij}\left|\mathrm{Phys}\right\rangle &= 0
\end{align}
\end{subequations}
\begin{subequations}
\label{lin_adm_grav_transform}
\begin{align}
\hat{A}_{ij} & \rightarrow \hat{A}_{ij} + \partial_{(i} \lambda_{j)}\\
\hat{E}_{ij} & \rightarrow \hat{E}_{ij} + \left(\delta_{ij}\partial^2 - \partial_i\partial_j\right)\phi
\end{align}
\end{subequations}
where Latin indices run over space while Greek run over spacetime, and $S_{(ij)}$ denotes symmetrization.

The local constraint Eq. \ref{lin_adm_grav_constraint}a is actually three constraints, one for each of the three components labeled by $j$.  These are the zero-momentum constraints on the ground state.  Starting instead from the local $SU(2)$ invariance of the frame fields and linearizing, one might conclude that it is a $U(1) \times U(1) \times U(1)$ gauge theory \cite{varadarajan2002}.  This is correct, but the three $U(1)$'s are not independent - they rotate into each other under a spatial rotation.  This follows from the fact that the charge $\rho_j$ is a vector, though care must be taken when making this identification\footnote{Formally this amounts to promoting the crystal symmetries to the rotation group in three dimensions.}.  This holds even if the background is not flat because the gauge constraint is by definition local.

We see that the curvature tensor is just
\begin{equation}
\hat{R}_{ij} = \epsilon_{iab}\epsilon_{jcd}\partial_a\partial_c \hat{A}_{bd}
\end{equation}
and so we are able to write the low-energy effective Lagrangian (after enforcing the constraints) as 
\begin{equation}
\mathcal{L} = E_{ij}\dot{A}_{ij} - \frac{J}{2}\left(E_{ij}^2 - \frac{1}{2}E_{ii}^2\right) - \frac{g}{2}A_{ij}R_{ij}.
\end{equation}
This is the Lagrangian for a spin-2 linearly-dispersing excitation, which we will call a graviton.  It can be shown to arise from a purely local bosonic lattice Hamiltonian \cite{0606100}, and the couplings $J$ and $g$ depend on the microscopics.  Much like the lattice model and corresponding field theory for electromagnetism, this model for linearized ADM gravity exists at an exactly stable IR fixed point, provided that the low-energy subspace enforces the gauge constraints.  We note in passing that this model appears to have a Chern--Simons-like term $\hat{A}_{ij}\hat{R}_{ij}$ which is only gauge-invariant up to boundary terms, but it will not modify the $3+1d$ topological properties of the model.

Gapless topological order is present in linearized ADM gravity, and the argument follows precisely as in QED.  If the system exists on $T^3$, one can thread a charge $\hat{\rho}_j$ around a non-contractible loop and annihilate it with an anticharge.  This leaves a flux of $\hat{E}_{ij}$ around the loop, which has energy density scaling as $1/L$.  As before, the flux is perpendicular to the surface $\Sigma$ and in the direction of $i$.  The new flux integrals are
\begin{equation}
\hat{\Phi}_{ij} = \int_\Sigma{dS_i \hat{E}_{ij}},
\end{equation}
where once more there is no summation over $i$.
These commute with each other and with the Hamiltonian, so we see that they characterize the gapless topological order of the ground state.  Moreover, there are three such fluxes for each surface - in the definition above, the surface is defined by the vector index $i$ while the index $j$ is free.  Thus, there are nine integers that characterize the ground state in the ``electric'' sector.

There is also a contribution from the ``magnetic'' sector due to the compactness of the gauge field $\hat{A}$; however, it is unimportant to the analysis as the electric sector already guarantees a degeneracy.  Moreover, the physicality of such linearized metric monopoles is difficult to justify in the full continuum theory.

\subsection{Entanglement Entropy}

A final point worth noting pertains to the entanglement structure of these theories.
In gapped systems, there is a universal constant term in the entanglement entropy across an arbitrary cut through the system.
This term characterizes the topological order \cite{kitaev2006}, and is constant because it reflects the charge-winding freedom and is hence independent of system size.

In gapless systems by contrast there is instead a universal coefficient of a (subleading) logarithmic term \cite{PhysRevB.94.125112, Radičević2016, 1112.5166, 1510.03863} due to both the (gapless) topological order and the photon.
This topological piece can be derived using the Bisognano-Wichmann theorem and charge conservation on the entanglement cut.
Intuitively the logarithmic scaling in system size arises because the spectrum near the ground state sector consists of a power-law of states, and so below any given cutoff the number of accessible states is a power-law.
The entropy is just the logarithm of that and hence is logarithmic in system size.

In the electromagnetic case the entanglement entropy, including the non-universal area-law part, is
\begin{equation}
S^{U(1)} = \alpha L^{d-1} + \left(\gamma^{U(1)}_{top} + \gamma^{U(1)}_{photon}\right)\log L
\end{equation}
where $\gamma^{U(1)}_{top} = (d-1)/2$ for space dimension $d$. 
Likewise in the gravitational case the arguments in \cite{PhysRevB.94.125112} permit us to calculate the universal coefficient of the $\log L$ term coming from topological order.  Since the charge is a $d$-dimensional vector and each component is independently conserved, one finds that
\begin{equation}
\gamma^{LG}_{top} = \frac{d(d-1)}{2}
\end{equation}
which gives $\gamma^{LG}_{top} = 3$ in $3+1d$.

The similarity in entanglement entropy between the electromagnetic and gravitational cases is striking, as is their difference from the case of gapped topological order.
This makes it clear that the phenomenon of gapless topological order is universal in the systems where it appears and simultaneously quite distinct from the more common notion of gapped topological order.

\section{Topological Degeneracy, Winding Operations, and Soft Bosons}
Now that we have stable lattice gauge theories with exactly gapless bosons, we want to consider the continuum limit.  We argue that the fundamental objects in these theories -- local constraints, gauge transformations, and global flux integrals -- carry over into the full continuum theory of electromagnetism and linearized gravity.  It is important to note that these connections are all made in the IR, where we expect the gauge constraints to hold - this is \textit{not} an attempt to build a full quantum theory of gravity.

The IR stability of these gauge theories follows from the local constraint on the low-energy Hilbert space.  For both of these systems (and the infinite family described in \cite{1601.08235}), this constraint is the conservation of some tensor-valued gauge charge.

In the previous section we followed the standard arguments to construct the degenerate ground states of QED and linearized gravity on the torus by starting with the lattice models and explicitly calculating the flux integrals.  Importantly, the states with nonzero flux are only degenerate in the limit of infinite system size, as the energies only go to zero as $1/L$.

However, the argument in that section depends on the topology in an awkward way, by relying on the periodicity of the perpendicular directions to the flux.  For concreteness, we calculate the commutator of $\hat{\Phi}_z$ with the (continuum) QED Hamiltonian along the surface $z=0$, which gives

\begin{equation}
\left[\hat{\Phi}_z, \hat{H}\right] = 2i\int{dxdy\ \epsilon_{zij} \partial_i \hat{B}_j} = 2i\int{dxdy \left(\nabla \times \hat{\boldsymbol{B}}\right)_z}.
\end{equation}
Provided that $\hat{\boldsymbol{B}}$ satisfies the periodic boundary conditions
\begin{equation}
\hat{\boldsymbol{B}}\left(\frac{L}{2},y,0\right) = \hat{\boldsymbol{B}}\left(\frac{-L}{2},y,0\right)
\end{equation}
and similarly for $y$, then the integral vanishes for even finite $L$.  The generalization to linearized gravity is straightforward.

Due to the reliance on the periodic boundary conditions in the perpendicular directions, this method is unsuited to showing the existence of the topological degeneracy in the more general case with just one periodic direction.  We are still able to construct the appropriate degenerate ground states however by using a winding construction adapted from the Minkowski spacetime arguments in \cite{1506.02906, 1601.00921, He2015} and similar arguments about lattice $SU(3)$ in \cite{PhysRevD.77.045013}.

\subsection{Winding}

\begin{figure}

%

\includegraphics{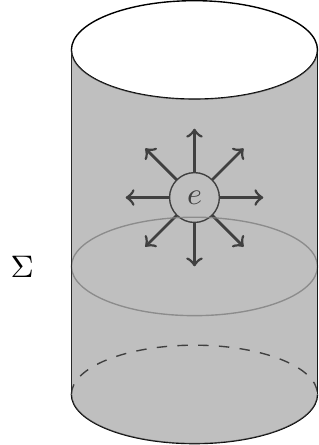}

\caption{A cylindrical system is shown with one possible cut surface $\Sigma$. The charges of interest are integrals over this surface of the normal component of the electric field.}
\label{fig:cutcylinder}
\end{figure}

\begin{figure}

%

\includegraphics{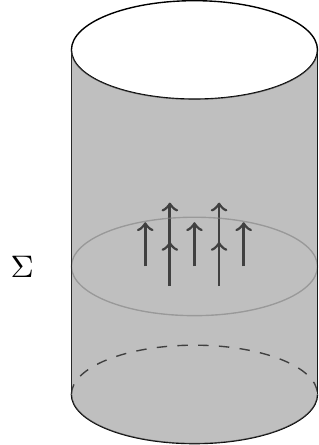}

\caption{A cylindrical system is shown with one possible cut surface $\Sigma$. The charges of interest are integrals over this surface of the normal component of the electric field. In this case the hard charges $\pm e$ (not shown) have been placed at distant mirrored positions on either side of the surface such that the field here is uniform and normal to the surface.}
\label{fig:cutcylinder3}
\end{figure}

We begin with open boundary conditions and consider a point charge $e$ located at $\boldsymbol{r}_0$.
This produces an electric field
\begin{equation}
	\boldsymbol{E} = \frac{e}{4\pi} \frac{\boldsymbol{r} - \boldsymbol{r}_0}{|\boldsymbol{r} - \boldsymbol{r}_0|^3}.
\end{equation}
If we partition the space with a planar surface $\Sigma$ as shown in Figure\ \ref{fig:cutcylinder}, the integral over this surface is easy to evaluate and yields
\begin{equation}
		\int_{\Sigma} \boldsymbol{E}\cdot d\boldsymbol{\Sigma} = \pm \frac{e}{2}\int_0^\infty \frac{r h}{(h^2+r^2)^{3/2}}dr = \pm \frac{e}{2},
\end{equation}
where $h$ is the distance from the charge to the surface, $r$ is the distance along the surface, and the sign of the integral depends on the sense of orientation of the surface.
If we now place a second charge $-e$ at $\boldsymbol{r}_1$ on the opposite side of the surface we find
\begin{equation}
		\int_{\Sigma} \boldsymbol{E}\cdot d\boldsymbol{\Sigma} = \pm e.
\end{equation}
Note that this result is independent of where we place the second charge, and so this integral only tells us about the total partition of charges across $\Sigma$.
As such we are free to move both charges as far away from the surface as we wish, leaving a field which is asymptotically constant, as shown in Figure~\ref{fig:cutcylinder3}.

\begin{figure}

\includegraphics{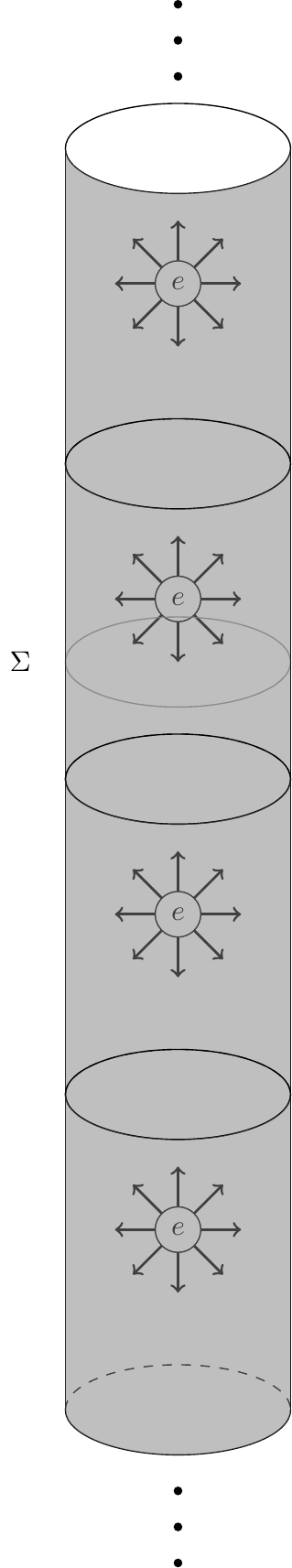}

\caption{A periodic unfolding of the system shown in Figure\ \ref{fig:cutcylinder}. The system is tiled a total of $N$ times but only the four closest to the surface $\Sigma$ are shown.}
\label{fig:cutcylinder2}
\end{figure}

Now suppose that we wish to impose periodic boundary conditions.
This may be done by ``unfolding'' the space and inserting periodically spaced copies of all charges, as shown in Figure\ \ref{fig:cutcylinder2}.
Of course this must be regularised when the system is finite, but in the limit as the system becomes infinite this procedure is correct.
If we place $N$ such pairs of charges, one from each pair on each side of the surface, the flux integral reads
\begin{equation}
		\int_{\Sigma} \boldsymbol{E}\cdot d\boldsymbol{\Sigma} = \pm N e.
\end{equation}
By contrast suppose we begin by placing a pair of charges at their periodic locations, but both on one side of the surface.
The integral will now vanish.
No matter how many times we do this, the integral still vanishes.
In the limit as the space becomes infinite this unfolding procedure remains perfectly well-defined, but the value of this global integral may be made to be any even integer simply by appropriate choice of the order in which it occurs.
Thus while local observable like the electric field converge by this process, the integral is sensitive to the order in which we place charges and hence the physical manner in which the periodic limit is reached.

The dependence of flux integrals on the manner in which we unfold the space corresponds precisely to the topological degeneracy in the theory.
This is because altering the order of placement corresponds in the periodic case to creating a dipole and winding it around the periodic dimension before destroying it.
To show this, we consider Poisson's equation for our pair of point charges:
\begin{equation}
	\nabla^2\phi = e\left(\delta(\boldsymbol{r} - \boldsymbol{r}_0) - \delta(\boldsymbol{r} - \boldsymbol{r}_1)\right).
\end{equation}
In momentum space this is
\begin{equation}
	-k^2 \tilde{\phi} = e\left(e^{i\boldsymbol{k}\cdot\boldsymbol{r}_0} - e^{i\boldsymbol{k}\cdot\boldsymbol{r}_1}\right).
\end{equation}
As a result
\begin{align}
	\tilde{\phi} = -\frac{e}{k^2}\left(e^{i\boldsymbol{k}\cdot\boldsymbol{r}_0} - e^{i\boldsymbol{k}\cdot\boldsymbol{r}_1}\right),
\end{align}
so
\begin{align}
	\tilde{\boldsymbol{E}} = ie\frac{\boldsymbol{k}}{k^2}\left(e^{i\boldsymbol{k}\cdot\boldsymbol{r}_0} - e^{i\boldsymbol{k}\cdot\boldsymbol{r}_1}\right)
	\label{eq:field}
\end{align}
The flux integral in momentum space is then
\begin{equation}
	\int_{\Sigma} \boldsymbol{E}\cdot d\boldsymbol{\Sigma} = -\int_{\Sigma}\int \frac{d^3\boldsymbol{k}}{(2\pi)^3}\left(-i\boldsymbol{k}\cdot\hat{n}\right)e^{-i\boldsymbol{k}\cdot\boldsymbol{r}} \tilde{\phi} d^2\boldsymbol{x},
	\label{eq:flux}
\end{equation}
where $\hat{n}$ is the unit vector normal to $\Sigma$.
For simplicity we may take the two dimensions parallel to $\Sigma$ to be infinite, in which case
\begin{equation}
	\int_{\Sigma} \boldsymbol{E}\cdot d\boldsymbol{\Sigma} = -\frac{i e}{2\pi}\int \frac{d k_n}{k_n}\left(e^{ik_n (r_{0,n} - r_{n})}-e^{ik_n (r_{1,n} - r_{n})}\right),
\end{equation}
where the subscript $n$ denotes the component normal to $\Sigma$.
Now if the remaining direction is periodic with finite size then the integral is actually a sum:
\begin{align}
	\int_{\Sigma} \boldsymbol{E}\cdot d\boldsymbol{\Sigma} &= -e\sum_{l=1}^\infty \frac{i}{2\pi l}\left(e^{2\pi i l (r_{0,n} - r_{n})/L}-e^{2\pi i l (r_{1,n} - r_{n})/L}\right)\nonumber\\
	 &- i\frac{e}{L}\lim_{k\rightarrow 0} \frac{e^{ik (r_{0,n}-r_n)}-e^{ik (r_{1,n}-r_n)}}{k}\\
	&= -e\sum_{l=1}^\infty \frac{i}{2\pi l}\left(e^{2\pi i l (r_{0,n} - r_{n})/L}-e^{2\pi i l (r_{1,n} - r_{n})/L}\right)\nonumber\\
	 & +e\left(\frac{r_{0,n}-r_{1,n}}{L}\right).
\end{align}
The special case-handling for the $l=0$ mode is necessary because this mode is degenerate in equation\ \eqref{eq:field}.
While any value will satisfy this component when $l=0$ (and hence $k=0$), we choose the value which is consistent with the limit as $k\rightarrow 0$, such that it remains well-defined and consistent with the integral formulation in the the limit as $L\rightarrow \infty$.
Now if we pick $r_{0,n} = -r_{1,n} = h$ and $r_n=0$, which we can do just by choice of $\Sigma$, then
\begin{equation}
	\int_{\Sigma} \boldsymbol{E}\cdot d\boldsymbol{\Sigma} = e\left[2\frac{h}{L}+\sum_{l=1}^\infty \frac{1}{\pi l}\sin\left(2\pi l \frac{h}{L}\right)\right].
\end{equation}
This may be evaluated as
\begin{equation}
	\int_{\Sigma} \boldsymbol{E}\cdot d\boldsymbol{\Sigma} = e\left[2\frac{h}{L} + \frac{i}{\pi}\log \frac{1-e^{2 i \pi  h/L}}{1-e^{-2 i \pi  h/L}}\right] = e\left[1+2\lfloor \frac{h}{L}\rfloor\right].
\end{equation}
As $h$ increases the charges wind around the torus, and as this happens the flux integral increments.
The offset of $1$ just comes from our choice of coordinates.
Note that the same argument holds when the remaining dimensions are finite.

\begin{figure}
\includegraphics[width=0.5\textwidth]{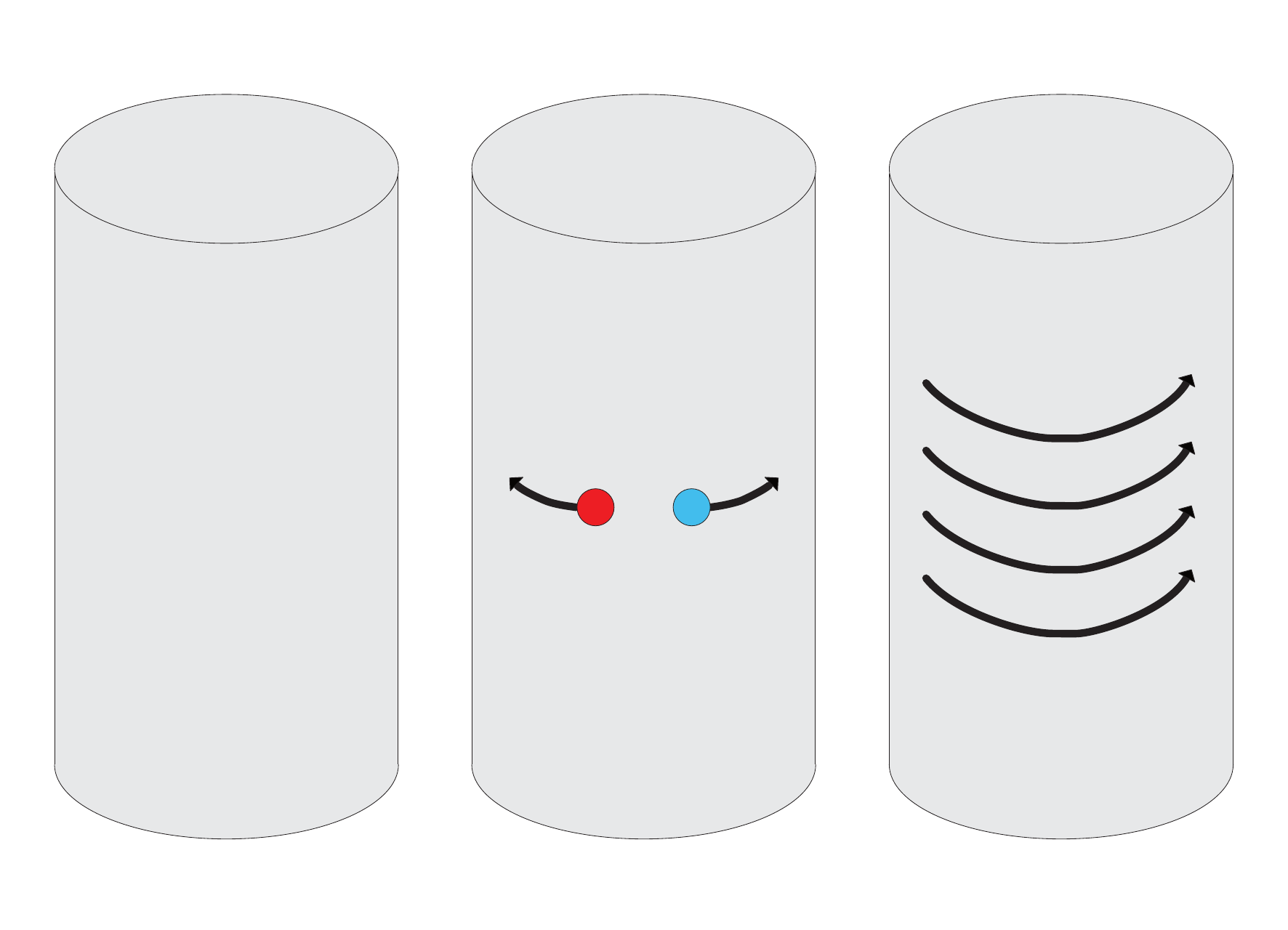}
\label{fig:winding}
\caption{Winding charges around a cylinder leaves a static uniform electric field pointing along the winding direction. For clarity the third spatial dimension is not shown.}
\end{figure}

The flux increment we see is associated with a mode with $k=0$, which is the soft photon sector.
In the high-energy context soft photons really are the vanishing-energy analogues of photons, but in the condensed matter language this is a bit of a misnomer, as it is not a photon mode but rather a large gauge transformation of the electric field.
To see that the mode really is soft note that the field associated with it is a static one which, when integrated over a surface of size $L^2$ yields a constant value.
That means the field scales as $L^{-2}$ and so the field energy density scales as $L^{-4}$.
Integrating over the volume gives energy scaling as $L^{-1}$.
As $L\rightarrow\infty$ this vanishes and so the modes associated with this winding procedure are actually soft.

This winding argument clearly holds for each periodic direction, and so on $T^3$ we find three independent integer-valued topological charges.
In a more complicated topology the number may vary.
For instance, consider a sphere with a hollowed-out center, and identify points on the outer edge with points at the same angular coordinates on the inner edge.
In this case the number of periodic directions scales as $L^2$, normalized by the UV lattice spacing.
These directions may be distinguished by the flux integral
\begin{equation}
	Q_\epsilon = \int_{\Sigma} \epsilon(\boldsymbol{r})\boldsymbol{E}\cdot d\boldsymbol{\Sigma}.
\end{equation}
By appropriate choice of $\epsilon$ this charge may be made sensitive to different winding directions $\hat{n}$.
This just alters the modes which are selected in integrating over the surface.
In this way we can decode the precise direction of each winding which has occurred.

This curious physics is strongly dimension-dependent.
To understand this note that in general a theory with $d$ spatial dimensions obeying a local flux constraint has field quanta with amplitude
\begin{equation}
	\psi \sim \frac{1}{L^{d-1}}.	
\end{equation}
This is just because the flux integral over a hypersurface of area $L^{d-1}$ must be independent of $L$.
The energy density is then
\begin{equation}
	\frac{dE}{dV} \sim \psi^2 \sim \frac{1}{L^{2d-2}}.	
\end{equation}
As a result the energy of the mode is
\begin{equation}
	E \sim L^d \frac{dE}{dV} \sim \frac{1}{L^{d-2}}.	
\end{equation}
In our universe, where $d=3$, this yields soft modes with energy scaling as $1/L$.
More generally $d=2$ is the critical dimension where the modes take on a constant energy independent of $L$.
Below this the modes are infinitely gapped in the thermodynamic limit and so are irrelevant.

\subsection{Soft Bosons and Topological Sectors}
We now explicitly connect the soft theorems to topological ground state degeneracy.
This has already been done in Minkowski spacetime \cite{1506.02906, 1601.00921, He2015} by proving that the Ward identities for the operators that detect topological sectors are equivalent to the soft photon and graviton theorems, though the degeneracy was not noted as topological in these works.
As a result we only need to show that these arguments continue to hold in the condensed matter language.

First we note that the equivalence described in \cite{1506.02906} follows from calculating the Liénard-–Wiechert fields for a massive particle-antiparticle pair and examining the field behavior near null infinity.  Due to the periodic boundary conditions placed on the  gauge fields at null infinity, this process can be viewed as the analogue of the winding procedures described above. However, instead of leaving the massive particles at $\mathcal{I}^{\pm}_{\mp}$, in the condensed matter system these particles are annihilated.

Since the particles have annihilated, they no longer contribute to the electric flux integrals that distinguish the topological degeneracy.  In the language of \cite{1506.02906}, there is no ``hard'' charge, and the only remaining piece is the ``soft'' charge left over from the winding procedure.  However, this ``soft'' charge encodes the history of the winding process for a given periodic direction, and is identified with the threaded electric flux.  This is the analogue of $\boldsymbol{\beta}$ being detectable in \citet{1601.00921}.

To make the connection between this winding process and the soft theorems explicit we must quantize the electric field, construct the soft photon operator corresponding to this winding procedure, demonstrate that its flux through $\Sigma$ matches that above, and show that it commutes with the Hamiltonian.
We begin by writing
\begin{align}
	\hat{\boldsymbol{A}} = \int \frac{d^3\boldsymbol{k}}{(2\pi)^3} \hat{a}_{i,\boldsymbol{k}} \boldsymbol{e}_i e^{i\omega t-i\boldsymbol{k}\cdot\boldsymbol{r}} + \mathrm{h.c.},
\end{align}
where $i$ is summed over, $\boldsymbol{e}_i$ form a basis of unit vectors and $\omega=k$ with the appropriate choice of units.
This allows us to write
\begin{align}
	\hat{\boldsymbol{E}} = \partial_t \hat{\boldsymbol{A}} = \int \frac{d^3\boldsymbol{k}}{(2\pi)^3} i\omega \hat{a}_{i,\boldsymbol{k}} \boldsymbol{e}_i e^{i\omega t-i\boldsymbol{k}\cdot\boldsymbol{r}} + \mathrm{h.c.}\\
\intertext{and}
	\hat{\boldsymbol{B}} = \nabla\times \hat{\boldsymbol{A}} = \int \frac{d^3\boldsymbol{k}}{(2\pi)^3} \hat{a}_{i,\boldsymbol{k}} \boldsymbol{k}\times\boldsymbol{e}_i e^{i\omega t-i\boldsymbol{k}\cdot\boldsymbol{r}} + \mathrm{h.c.}.
\end{align}

We now wish to construct the soft photon operator $W_{\hat{n}}$ which produces the field associated with winding a pair of charges around a periodic dimension of the system.
As this is a static field it is described by a coherent state.
This means that the operator which creates it is a displacement operator, so
\begin{align}
	\hat{W}_{\hat{n}}^\dagger(h) &= \exp\left(e \int \frac{d^3\boldsymbol{k}}{(2\pi)^3}\frac{e^{i\boldsymbol{k}\cdot\hat{n}h}-e^{-i\boldsymbol{k}\cdot\hat{n}h}}{k^2}\hat{a}_{\hat{k},\boldsymbol{k}}^\dagger + \mathrm{h.c.}\right).
\end{align}
This indeed produces the field in equation\ \eqref{eq:field}, as
\begin{align}
	\langle 0| \hat{W}_{\hat{n}}(h) \hat{\boldsymbol{E}} \hat{W}_{\hat{n}}^\dagger(h)|0\rangle &= e\int \frac{d^3\boldsymbol{k}}{(2\pi)^3}i\omega \frac{\boldsymbol{k}}{k} \frac{e^{i\boldsymbol{k}\cdot\hat{n}h}-e^{-i\boldsymbol{k}\cdot\hat{n}h}}{k^2}\\
	&= ie\int \frac{d^3\boldsymbol{k}}{(2\pi)^3}\boldsymbol{k} \frac{e^{i\boldsymbol{k}\cdot\hat{n}h}-e^{-i\boldsymbol{k}\cdot\hat{n}h}}{k^2},
\end{align}
where $|0\rangle$ is the vacuum state annihilated by $|a_{i,\boldsymbol{k}}\rangle$.
It follows that the flux $\hat{W}_{\hat{n}}$ carries across $\Sigma$ is the same as the classical flux in equation\ \eqref{eq:flux} when $h = L$, so this operator does in fact correspond to the winding process.

Finally to see that $\hat{W}_{\hat{n}}(L)$ is indeed a soft operator note that the Hamiltonian is given by
\begin{align}
	\hat{H} = \int d^3\boldsymbol{r} |\hat{\boldsymbol{E}}|^2 + |\hat{\boldsymbol{B}}|^2.
\end{align}
The magnetic component vanishes because $\boldsymbol{A} \parallel \boldsymbol{k}$ and $\boldsymbol{B}\propto \boldsymbol{k}\times\boldsymbol{A}$.
The electric component may be resolved in Fourier space as
\begin{align}
	\int d^3 \boldsymbol{r}|\hat{\boldsymbol{E}}|^2 = \int \frac{d^3\boldsymbol{k}}{(2\pi)^3} k^2 \sum_{\hat{m}} \hat{a}_{\hat{m},\boldsymbol{k}}^\dagger \hat{a}_{\hat{m},\boldsymbol{k}},
\end{align}
where $\hat{m}$ range over an orthonormal basis.
Now note that
\begin{align}
	\hat{a} e^{\alpha \hat{a}^\dagger - \alpha^* \hat{a}} = e^{-\alpha \hat{a}^\dagger + \alpha^* \hat{a}}(\hat{a} + \alpha)
\end{align}
so
\begin{align}
	\left[\hat{a}^\dagger \hat{a}, e^{-\alpha \hat{a}^\dagger + \alpha^* \hat{a}}\right] = e^{-\alpha \hat{a}^\dagger + \alpha^* \hat{a}}\left(|\alpha|^2 + \alpha \hat{a}^\dagger + \alpha^* \hat{a}\right).
\end{align}
As a result
\begin{align}
	\left[\hat{H},\hat{W}_{\hat{n}}(L)\right] &= \hat{W}_{\hat{n}} \int \frac{d^3\boldsymbol{k}}{(2\pi)^3}k^2 e^2\left|\frac{e^{i\boldsymbol{k}\cdot\hat{n}L}-e^{-i\boldsymbol{k}\cdot\hat{n}L}}{k^2}\right|^2\nonumber\\
	&+ek^2 \left(\frac{e^{i\boldsymbol{k}\cdot\hat{n}L}-e^{-i\boldsymbol{k}\cdot\hat{n}L}}{k^2}\left(a_{\hat{k},\boldsymbol{k}}^\dagger - a_{\hat{k},\boldsymbol{k}}\right)\right)\\
	&= \hat{W}_{\hat{n}} \int \frac{d^3\boldsymbol{k}}{(2\pi)^3} 4e^2\frac{\sin^2(k_n L)}{k^2}\nonumber\\
	 &+ 2i e \sin(k_n L)\left(\hat{a}_{\hat{k},\boldsymbol{k}}^\dagger - \hat{a}_{\hat{k},\boldsymbol{k}}\right).
	 \end{align}
If the system is periodic along $\hat{n}$ then the integral in that dimension must be replaced by a sum so
\begin{align}
		\left[\hat{H},\hat{W}_{\hat{n}}(L)\right]&= \frac{1}{L}\hat{W}_{\hat{n}} \sum_{l=0}^{\infty}\int \frac{d^2\boldsymbol{k}_{\perp}}{(2\pi)^3} 4e^2\frac{\sin^2(2\pi l)}{k^2}\nonumber\\
	 &+ 2i e \sin(2\pi l)\left(\hat{a}_{\hat{k},\boldsymbol{k}}^\dagger - \hat{a}_{\hat{k},\boldsymbol{k}}\right).
\end{align}
In this form it is clear that all modes with $l\neq 0$ vanish.
The second term vanishes when $l=0$ but the first does not.
In particular if $k_{\perp} = 0$ as well then the first term does not vanish.
As a result the only contribution comes from the term with $k=0$.
This term must be written as a limit in order to ensure continuity in the vicinity of $h=L$, and this limit must be approached along the $\hat{n}$ direction, as the field is in this direction when $h=L$.
Thus we find
\begin{align}
	\left[\hat{H},\hat{W}_{\hat{n}}(L)\right]&= \frac{4e^2}{L^3}\hat{W}_{\hat{n}} \lim_{k\rightarrow 0} \frac{\sin^2(k_n L)}{k_n^2}\\
	&= \frac{4e^2}{L}\hat{W}_{\hat{n}} \lim_{u\rightarrow 0} \frac{\sin^2(u_n)}{u_n^2}\\
	&= \frac{4e^2}{L}\hat{W}_{\hat{n}}.
\end{align}
As promised this vanishes as $L^{-1}$ and so the operator is indeed soft.

Thus we have shown that the operation which winds a particle-antiparticle pair around a large loop has support preferentially as $k\rightarrow 0$, and scales in such a way that it commutes with the Hamiltonian up to terms of order $L^{-1}$. As such we identify it as the soft photon operator in our system. 

\subsection{Local Indistinguishability and Generalizations}

Now that we have shown the ground state degeneracy on a torus, we turn to another important characteristic of topological order - local degeneracy.  Roughly speaking, this means that any local measurement should be unable to determine which topological sector the system occupies.  We can see this heuristically by noting physical processes with typical length scale $\Delta X$ cannot resolve momenta more precisely than $\Delta P \sim \Delta X^{-1}$.  Since the topological degeneracy comes from the $1/L$ modes, local measurements with $\Delta X \ll L$ cannot determine the topological sector.

Though a more complete field-theoretic treatment is left to future work, we can get some intuition about this result by considering the careful treatment of IR divergences first discussed in \cite{PhysRev.140.B516}.  By considering scattering processes that both involve virtual infrared bosons and the emission/absorption of infrared bosons, the IR divergence goes away.  The new transition rate is given in terms of a positive function $C$ that depends on the details of the gauge theory, a positive function $b$, the UV cutoff $\Lambda$, the total energy of emitted soft modes $E$, and the original transition rate $\Gamma_{\alpha\beta}^0$:
\begin{equation}
\Gamma_{\alpha\beta} = (E/\Lambda)^C b(C) \Gamma_{\alpha\beta}^0
\end{equation}
When considering the modes that change topological sectors, we see that the energy goes to zero as $1/L$ and thus the transition rate for local scattering processes to change topological sectors vanishes.  In this light, the IR divergence in QED and linearized gravity is similar to IR divergences in spontaneously broken (0-form) symmetries.
This is in agreement with the arguments of~\citet{0264-9381-34-20-204001}, though it does not mean as has been claimed~\citep{PhysRevD.97.046014} that such modes are unmeasurable or trivially decoupled, simply that modes at some asymptotic distance $L$ require space and time proportional to $L$ to measure.

This identification solves the longstanding question of the connection between $1/L$ photon modes and topological sectors.  That is, $1/L$ photon modes carry the charges which identify different topological sectors.  This is a significant point, but in retrospect is not entirely surprising, as the $1/L$ modes by definition are sensitive to global physics.

The arguments above do not rely on any details of QED other than charge conservation and the existence of gapless modes, which combined allows us to determine the scaling of the fields.
The general nature of this construction then leads to the somewhat remarkable conjecture that any stable, deconfined, continuous gauge theory with a soft theorem should have some notion of topological degeneracy.  The different topological sectors can be reached by winding gauge-charged matter around the large loops of the torus, which can be interpreted as threading (locally invisible) soft bosons. Importantly, one only expects stability provided that the matter is massive, so that there is an exponential cost to ``unwind'' the topological sectors.

Such behavior does not extend to gapless theories without a gauge structure (such as superfluids), since there are no electric fields or charges, nor large gauge transformations.  From the viewpoint of \cite{1506.02906}, there are no hard charges and thus the Ward identity is not equivalent to a flux integral.  It should be noted that while there is a notion of ground-state degeneracy in systems with spontaneously broken continuous symmetries, this degeneracy is not dependent on the topology of the manifold on which the system resides.

\subsection{Higher Form Symmetries}
\label{sec:sym}
The process of creating a charge-anticharge pair and moving them around a closed path $C$ is a well-known object in gauge theories: Wilson loops.  Confinement of the gauge theory can be determined by the area or perimeter law scaling of these objects, captured in the Wilson loop operator

\begin{equation}
W = e^{\int_C A}
\end{equation}

This is manifestly invariant under the ordinary gauge transformation, since the integral of $df$ vanishes.  However, we could consider a more general gauge transformation:

\begin{equation}
A \rightarrow A + \lambda
\end{equation}

If we require that $d\lambda = 0$ but $\lambda \neq df$ for any $f$, then the field strength $F = dA$ is left invariant but the Wilson loops change by a factor of $\exp{\int_C \lambda}$.

This symmetry, known as a 1-form symmetry, is one of the infinite family of generalized global symmetries with very interesting properties\cite{gaiotto2015}.  As opposed to 0-form symmetries, whose charges are point-like objects such as particles, the 1-form symmetries act line-like objects.  The elements of the 1-form symmetries act on surfaces, which for $U(1)$ are

\begin{equation}
U(\alpha,M^2) = \exp\left(i\frac{\alpha}{2g^2}\int_{M^2}\star F\right)
\end{equation}

These operators form a 2-group due to the ways the manifolds can be stacked.  However, for fixed $M^2$ (say, the $xy$-plane) this symmetry is just a $U(1)$ symmetry, and its irreduciple representations are labeled by integers.  This symmetry can spontaneously break, giving rise to a Goldstone boson, namely the photon.  Charged matter explicitly breaks this symmetry, but for energies much smaller than the charge gap the symmetry is restored.  Whether or not the a symmetry spontaneously breaks determines whether or not the gauge theory is deconfined\cite{gaiotto2015}.

Our previous discussion can thus be rephrased in this language as identifying ``topological'' surfaces that live in the homology of the manifold, specifically closed surfaces that do not bound a volume, and noting that, in the thermodynamic limit, acting with the generators of the 1-form symmetry moves between ground states.  This unifies gapped and gapless topological order in gauge theories, in that they both have spontaneously broken 1-form symmetries.  In fact, the ground state degeneracy in $SU(3)$ noted in\cite{PhysRevD.77.045013} is topological in the same way.

A more complete analysis of higher-form symmetries and their relation to topological phases is left to future work.

\section{Topological Order in Open Systems}

In the previous two sections we argued that the IR fixed point for both electromagnetism and linearized gravity should have a well-defined continuum gauge theory arising from the local constraints of charge and momentum conservation.
These Gauss-law type constraints give rise to a ground-state degeneracy on a torus, which we have termed gapless topological order due to the presence of the gauge bosons.

More specifically, the generators of topological degeneracy in gravitation and electromagnetism are the charge operators of \citet{1601.00921}, written schematically as
\begin{equation}
\Hat{Q}^\pm_{\epsilon,\mathrm{EM}} \sim \int_{\mathcal{I}_{\mp}^{\pm}} \epsilon \star F
\label{eq:emCharge}
\end{equation}
for electromagnetism, with a similar integral over $\mathcal{I}$ or the equivalent boundary surface holding for gravitation.

These operators generate the ground-state degeneracy of the vacuum, such that
\begin{equation}
\langle 0 | \hat{Q} | 0 \rangle = 0
\label{eq:change}
\end{equation}
and
\begin{equation}
\left[\hat{H},\hat{Q}\right]=0.
\label{eq:commutator}
\end{equation}
Eq.\ \eqref{eq:commutator} follows because $\hat{Q}$ represents a soft mode corresponding to an asymptotic symmetry.
Eq.\ \eqref{eq:change} simply represents the fact that the field configuration associated with a system containing a soft boson is distinct from that of a system not containing it.
In a finite universe the commutator is of order $L^{-1}$, matching the condensed matter case.

\begin{figure}


\includegraphics{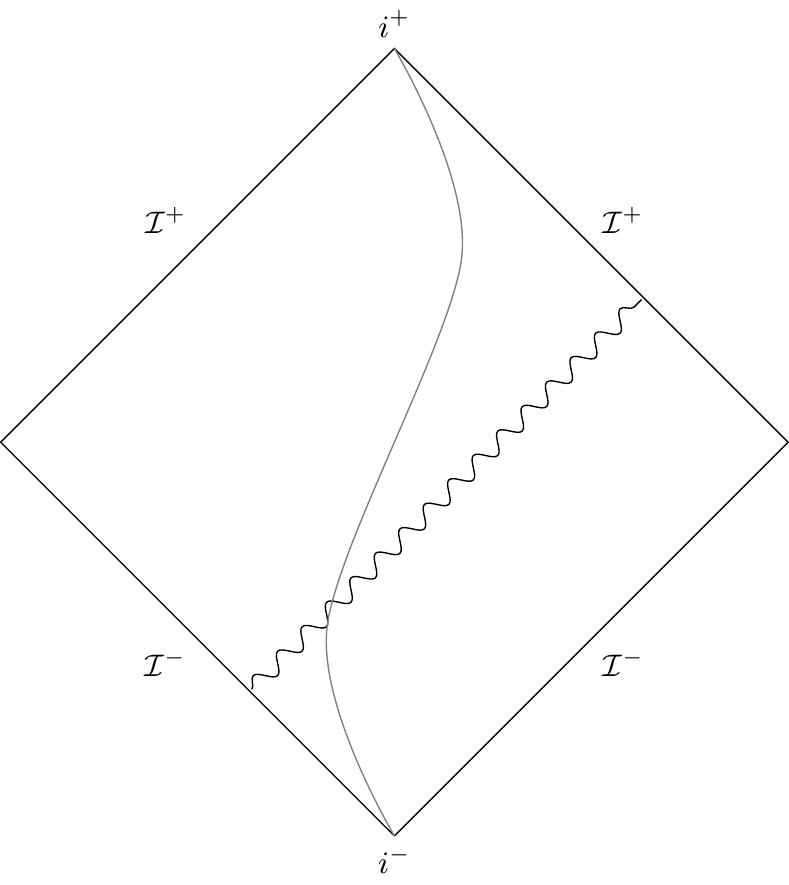}

\caption{Massless particles (wiggling lines) travel between past and future null infinity while massive ones (regular lines) travel between $i^-$ and $i^+$. This difference means that massive (gapped) charges accumulate at $i^{\pm}$ while soft charges appear at $\mathcal{I}^{\pm}_{\pm}$.}
\label{fig:penrose1}
\end{figure}

This topological degeneracy is a surprising result of the metric signature which holds even in open spacetimes.  
To see this, note that a key result enabling equations \eqref{eq:change} and \eqref{eq:commutator} is the antipodal mapping, which associates antipodal points on the boundaries of the past and future \cite{1601.00921}.
This mapping emerges because the ground state involves only massless modes, which propagate through the spacetime bulk at $c$.
More specifically, the antipodal mapping is possible because in the absence of charges massless fields are fully determined by their values on any one Cauchy surface, as the wave equation
\begin{equation}
	\square \phi = 0
	\label{eq:massless}
\end{equation}
may be used to propagate them from that surface to the rest of spacetime.
This establishes a correspondence between the values these fields take on at $\mathcal{I}^{-}$ and $\mathcal{I}^{+}$.
As a result we may write as a slight rephrasing of \citet{1601.00921}
\begin{equation}
	\phi^- \left(u\right) = e^{i\alpha(u)} \phi^+ \left(u\right)
	\label{eq:relationmassless}
\end{equation}
where $u$ is the relevant null coordinate, $\phi^{\pm}$ are evaluated at antipodal points on the corresponding null surfaces, and $\alpha$ is a function dependent on the gauge condition taken at these surfaces.

Upon threading a flux quantum through from one null surface to the other an overall factor of $e^{i\alpha}$ is accumulated.
This factor may be set to unity by appropriate choice of gauge to yield periodic boundary conditions \cite{1506.02906}.
Even without doing this it is clear that Eq.\ \eqref{eq:relationmassless} connects antipodal points on the space, and so in the asymptotic compactification gives it topological structure.
This is shown in Fig.\ \ref{fig:penrose1}.
This sidesteps the problem of the topology of the universe, since we need not specify the genus of spacetime. 

\begin{figure}

%
%

\includegraphics[width=0.45\textwidth]{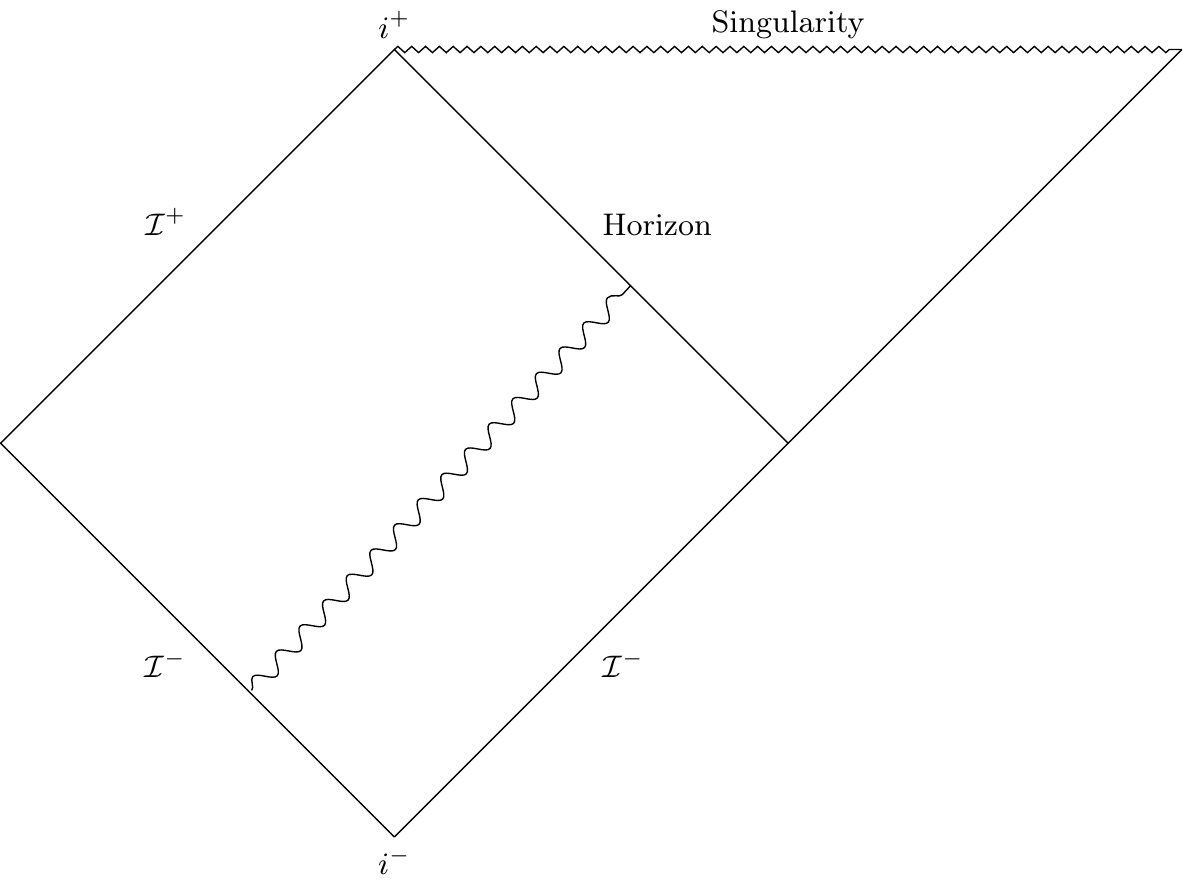}

\caption{Massless soft graviton (wiggling line) propagates from $\mathcal{I}^{-}$ and impinges on the black hole (top-right), resulting in entanglement between the two horizons.}
\label{fig:penrose2}
\end{figure}

In the presence of a more complicated topology or additional horizons (i.e. black holes) the identification is between modes on different horizons which overlap when propagated both forwards and backwards in time.
For a simple example, consider a soft graviton with large angular quantum numbers, such that it is highly directed.
This graviton propagates from a region on $\mathcal{I}^-$ until it encounters a black hole.
The graviton becomes bound to the event horizon by scattering into one of the surface soft modes.
This process is shown in Fig.\ \ref{fig:penrose2}.
An analogous equation to Eq.\ \eqref{eq:relationmassless} relates the mode which arrives in this fashion at $\mathcal{I}^{-}$ to the mode which arrives on the black hole's horizon.
More generally, if the mode did not fully scatter onto the black hole's horizon there would be a relation between the three boundaries, namely the black hole, $\mathcal{I}^-$, and $\mathcal{I}^+$, with at most a phase accumulation between each of them.

Massive charged excitations by definition propagate from $i^-$ to $i^+$ and break this structure by introducing scattering processes, but this propagation is exponentially suppressed as $e^{-m L}$ and hence does not break the ground state degeneracy in the thermodynamic limit.
This is the analogue of the circumferential create-wind-destroy propagation process on a torus.
A key difference is that topological winding in a $3+1d$ spacetime is complicated by its infinite nature, which makes it the case that winding soft flux through the universe requires an infinite amount of time, or at least the time required to reach an acceptable approximation of asymptotic infinity.

For a concrete example consider inserting a single $+e$ electric charge at $i^-$ with velocity $\boldsymbol{\beta}$.
The electromagnetic tensor at $\mathcal{I}^{\pm}_{\mp}$ is \citep{1601.00921}
\begin{equation}
	F_{\pm,rt} = \frac{e(1-\beta^2)}{4\pi r^2(1\mp\boldsymbol{\beta}\cdot\hat{r})^2},
\end{equation}
where the subscript $rt$ denotes the component which goes as $dr \wedge dt$.
The soft charge at $\mathcal{I}^{\pm}_{\mp}$ is then
\begin{align}
	Q_\epsilon^{\pm} &= \frac{1}{e^2}\int_{\mathcal{I^{\pm}_{\mp}}} \epsilon \star F_{\pm,rt}\\
		&= \frac{1}{e}\lim_{r\rightarrow \infty} r^2\int_{S^2} \frac{\epsilon(1-\beta^2)}{4\pi (1-\boldsymbol{\beta}\cdot\hat{r})r^2}\\
		&= \frac{1}{e}\int_{S^2} \frac{\epsilon(1-\beta^2)}{4\pi(1-\boldsymbol{\beta}\cdot\hat{r})}.
\end{align}
Now suppose that we thread a negative charge $-e$ with velocity $\boldsymbol{\beta'} \neq \boldsymbol{\beta}$.
The net charge is evidently
\begin{equation}
	Q_{\epsilon}^{\pm} = \frac{1}{e}\int_{S^2} \frac{\epsilon}{4\pi}\left[\frac{(1-\beta^2)}{1-\boldsymbol{\beta}\cdot\hat{r}}-\frac{(1-\beta'^2)}{1-\boldsymbol{\beta}'\cdot\hat{r}}\right].
\end{equation}
This is nonzero though the net charge which has been threaded through $i^{\pm}$ is zero.
Of course if $\beta \neq \beta'$ then the charges are always located far apart spatially at $i^{\pm}$, but in the compactified coordinates in which we identify $i^{-}$ with $i^+$ this is fine.
A similar argument holds for black hole event horizons and for supertranslations, but unfortunately the fields are much more difficult to write out explicitly.
The key difference is that the charges must begin at $i^{-}$, enter the black hole, be emitted via Hawking radiation, and then head towards $i^+$.
Other than that the argument is precisely the same, and the integral may be taken over the horizon of the black hole along with the necessary spatial cut to reach $\mathcal{I}^{\pm}_{\mp}$.

It is usually useful to think of topological winding as an operation which may be iterated.
This is not the case for the universe due to the infinite time required to wind.
It is worth asking then in what sense this order is topological.
The answer is twofold.
First, while we cannot iterate a winding process on a single universe, we can simulate the process given several spacetimes.
To see this suppose we start with a universe with no soft charge.
We can then wind a dipole from $i^-$ to $i^+$ as described above and note the soft charge which appears.
We can set up a second empty spacetime with this soft charge from the beginning.
If we then wind a dipole through that spacetime the soft charge doubles, and so it is clear that the amount of soft charge is a quantity which changes with dipole winding, which is locally unobservable, and which permits us to indefinitely move from sector to sector via this process.
These are the hallmarks of topological order.

The second argument is somewhat more direct: it is entirely valid to wind multiple dipoles simultaneously, as they can be separated spatially yet lead to the same soft charge so long as their asymptotic velocities are the same.
As a result it is sensible to talk about iterated winding, just with the iteration occurring in space rather than time.

What both of these arguments fail to address is what happens to the hard charge at infinity.
This likewise has two answers which differ just as a matter of interpretation.
First, suppose we bring a dipole out of the vacuum at some point near $i^-$ and then wind it to some point near $i^+$ before annihilating it.
Far from the origin we may draw a surface and integrate over it to measure the resulting soft charge.
This is the case so long as the creation and annihilation occur outside of the surface, and so the winding effectively encompasses a loop in spacetime between the creation and annihilation points.
The limit may then be taken as this surface goes off to infinity, keeping these two points outside as it goes.

The alternative interpretation of this process is that we may `glue' two spacetimes together as a result of the periodic boundary conditions at infinity.
In this process, $\mathcal{I}^+$ and $i^+$ in one spacetime are identified with the antipodal $\mathcal{I}^-$ and $i^-$ in the other spacetime, and vice-versa.
As a result a charge wound from $i^-$ to $i^+$ in one spacetime simply carries on to the next one, before wrapping back to the first spacetime once more.
In this way we avoid formal accumulation of charge at infinity.

Regardless of interpretation, it is clear that there is topological order in these systems, both as a result of the odd boundary conditions associated with an open spacetime and as the continuum limit of the corresponding lattice systems.
This order manifests via global operators that distinguish a charge which is not locally measurable, and which have a direct connection to the hard (local) charge wound through the system.

\section{Limits}

As the topological order discussed here is quite broad in nature it is worth discussing the limits in which it is applicable.
In particular it relies primarily on two key assumptions; linearity and low-energy (IR).

Linearity in this context does not mean that the metric is a small perturbation against Minkowski space.
Rather, it means that the quantum mechanical perturbations we consider correspond to small metric perturbations against whatever background metric we choose.
This is equivalent to saying that all perturbing gravitational waves have small amplitude, or equivalently that gravitons are not so prevalent as to interact strongly with one another.
In fact we do not even require that this be true universally, as we only need it to hold in the regions around which we perform flux integrals.
The gravitational field may be perturbed in an arbitrarily nonlinear manner outside of these regions, and these nonlinear effects will appear simply as fluxes of the relevant conserved charges through the bounding surface.

Along similar lines, working in the low-energy (IR) limit means that we are considering gravitons with energies of order $1/L$, where $L$ is a characteristic scale for the universe.
This is true even in the presence of a black hole, where the scale of the universe and not that of the black hole remains the relevant parameter.
This is because in an infinite universe, black holes support precise zero modes reflecting the BMS symmetry of relativity.
The fact that we confine our discussion to these modes does not make our conclusions any weaker, however, as our claim is precisely that these modes give rise to topological order.
The existence of higher-energy modes is irrelevant to this point.

\section{Black Holes and Information}

As mentioned previously, there are modes which exist on the event horizon of a black hole which are analogous to the modes on the horizons at infinity.
These modes actually obey the same dispersion relation up to local horizon distortions, just with the expansion coordinate converted from $\xi = 1/r$ to $\xi = r-r_s$, $r_s$ being the horizon radius\ \cite{1601.00921}.

Now consider the formation of a Hawking pair at the horizon.
For simplicity, we consider QED, so the state is a charge singlet of charge-1 particles.
The state is then one of the Bell states, given by
\begin{equation}
	|\phi\rangle = \frac{1}{\sqrt{2}}\left(|+-\rangle + |-+\rangle\right).
\end{equation}
One particle falls into the interior while the other escapes to $\mathcal{I}^{+}$.
By the preceding arguments there are flux integrals which can detect the fact that a particle has escaped.
These integrals measure the soft charge on the horizon, and so the state of the outgoing particle must be entangled with the soft sector.
If these integrals can detect the degree of freedom we have considered, the state must really be
\begin{equation}
	|\phi\rangle = \frac{1}{\sqrt{2}}\left(|+-+\rangle + |-+-\rangle\right)
	\label{eq:ghz}
\end{equation}
up to a minus sign and overall phase factor, where the additional qubit describes the state of the flux integral that labels the soft sector.
In this way it is possible to entangle the outgoing particle with the soft sector.
Now Eq.\ \eqref{eq:ghz} is the GHZ state for three particles, and so we know that if we trace out the soft sector there will be no remaining entanglement between the infalling and outgoing Hawking particles.
This ought to occur for all portions of the state of the particle which may be read from soft flux integrals, and so if these indeed encode all of the information which falls in then there is no firewall paradox.

\begin{figure}
\includegraphics[width=0.5\textwidth]{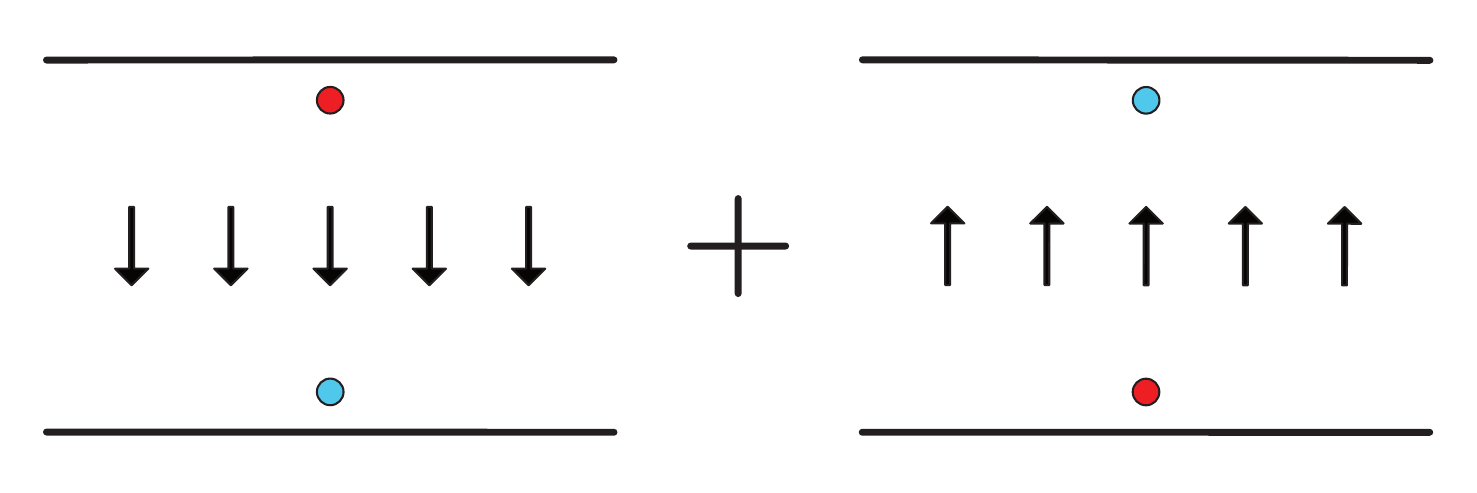}
\label{fig:winding2}
\caption{Separating charges in different directions generates a GHZ state between the charge positions and the gauge field.}
\end{figure}

This resolution amounts to quantum mechanical violation of equivalence via the monogamy of entanglement, and is essentially a physical realization of the nonlocal gravitational modes proposed by\ \citet{2016arXiv160704642O}.
Notably this exchange of entanglement is a purely quantum mechanical effect.
The soft theorems guarantee that interactions with the soft sector are not classically measurable, so this resolution of the paradox represents a way to preserve the classical equivalence principle while minimally violating it quantum mechanically.

It is important to emphasize that this argument does not resolve the broader information paradox.
To see this note that the Bell state is not recoverable from $|\phi\rangle$ after tracing out the particle which fell in.
This is another way of saying that the information is not transferred from the particle to the horizon nor is it cloned, it is just entangled with the horizon.

\section*{Conclusion}

We have argued that a peculiar type of gapless topological order exists in the lattice models of electromagnetism and linearized gravity, and that these models both flow to exactly stable IR fixed points with well-behaved continuum descriptions.  Thus, we can use this topological order to characterize the IR behavior and ground state of the continuum theories, provided that the gauge constraints hold (i.e. the metric deviations are small).

While there is no natural way to impose periodic boundary conditions on the universe, we have used the Lorentzian signature of the metric to identify non-contractible loops of the gauge fields in spacetime, allowing for the construction of non-local operators which commute with the Hamiltonian and whose eigenvalues distinguish the various ground states.  Finally, we have connected all of these objects to well-known results in the literature.

We have seen that gapless topological order, as described in this paper, shares many properties with ordinary topological order.  Primarily, they both have a family of locally indistinguishable ground states, and are degenerate on a torus.  However, we have made very general arguments for both the local indistinguishability and degeneracy, and thus expect the arguments to hold for other deconfined continuous gauge theories with soft boson theorems.  A precise characterization and proof is left to future work.

Finally, we have discussed applications of this work to black holes, with the key insight that the firewall paradox may be resolved by reducing the equivalence principle to be purely classical, with violation at the level of entanglement.
This is suggestive of a phase transition in the vacuum across the event horizon, but we leave a more detailed analysis of this phenomenon to a later work.

\section*{Acknowledgments}

The authors gratefully acknowledge numerous and significant conversations with Gil Refael.
The authors also thank C. Xu, D. Else, C. Vafa, A. C. Davies and Y. Ben Tov for many helpful discussions. A.D.R. acknowledges financial support from NSF Grant No. DMR-1151208, and thanks the Simons Center for Geometry and Physics for hospitality that aided this work. A.S.J. acknowledges financial support from a Marshall Scholarship as well as travel funds from a Hertz Fellowship which aided in this work.

\bibliography{refs}

\end{document}